\documentclass[twocolumn,showpacs,preprintnumbers,amsmath,amssymb,superscriptaddress, aps]{revtex4}

\usepackage{graphicx}% Include figure files
\usepackage{dcolumn}% Align table columns on decimal point
\usepackage{bm}% bold math

\begin{document}

\title{Phase diagram of Bose-Fermi mixtures in one-dimensional optical lattices}

\author{Lode Pollet}
%\email{pollet@itp.phys.ethz.ch}
 %\altaffiliation[Also at ]{Institut Theoretische Physik, ETH H\"onggerberg, CH-8093 Z\"urich \\ Switzerland}
\author{Matthias Troyer}%
\affiliation{Institut Theoretische Physik, ETH Z\"urich, CH-8093 Z\"urich, Switzerland}

\author{Kris Van Houcke}
\author{Stefan M. A. Rombouts}
\affiliation{Subatomaire en Stralingsfysica, Universiteit Gent, Proeftuinstraat 86, B - 9000 Gent, Belgium}

\date{\today}% It is always \today, today,

\begin{abstract}
The ground state phase diagram of the one-dimensional Bose-Fermi Hubbard model is studied in the canonical ensemble using a quantum Monte Carlo method. We focus on the case where both species have half filling in order to maximize the pairing correlations between the bosons and the fermions. In case of equal hopping we distinguish between phase separation, a Luttinger liquid phase and a phase characterized by strong singlet pairing between the species. True long-range density waves exist with unequal hopping amplitudes.
\end{abstract}

\maketitle

Ultracold atomic gases loaded in an optical lattice offer unique possibilities to simulate many fundamental models of solid state physics. The system is extremely clean and there is unprecedented experimental control over the lattice period and the interaction strength. The prediction~\cite{Jaksch98} and the successful observation of the bosonic superfluid-Mott transition by Greiner {\it et al.}~\cite{Greiner02} was the key breaktrough and led to a revival in interest in bosonic physics. Experiments on fermions in different hyperfine states are under way~\cite{Kohl05} and can potentially be employed for studies of high-Tc superconductivity and Mott-insulating phases. Ultracold atoms can also be applied to create a completely different and unique state of matter: bosonic and fermionic atoms can simultaneously be trapped in order to let them interact in a controlled way.
%Interest in ultracold atomic gases loaded in  an optical lattice was triggered by the prediction~\cite{Jaksch98} and the successful observation of the bosonic superfluid-Mott transition by Greiner {\it et al.}~\cite{Greiner02}. 
%Many fundamental models of solid state physics could be realized experimentally if the bosons could be replaced by fermions. Unfortunately, fermions are harder to cool beyond degeneracy. Progress was made using the attractive interaction between bosons and fermions, when the fermions cool by thermalization to the cooler bosonic cloud. 
%It was soon realized that such Bose-Fermi mixtures are worth studying as well. 
Nowadays, several stable Bose-Fermi mixtures have been created , e.g., $^7{\textrm {Li}}-{}^6{\textrm {Li}}$~\cite{Truscott01}, $^{23}{\textrm {Na}}-{}^6{\textrm {Li}}$~\cite{Hadzibabic03}, $^{87}{\textrm {Rb}}-{}^{40}{\textrm {K}}$~\cite{Roati03} and $^{87}{\textrm {Rb}}-{}^{6}{\textrm {Li}}$~\cite{Sibler05}, both in and without an optical lattice. These experimental advances have challenged theoreticians: the properties of mixtures have been studied in Refs.~\cite{Lewenstein04, Albus03,Fehrmann04, Cazalilla03, Mathey04, Imambekov05a, Imambekov05b, Roth03}, using a variety of methods. Most remarkably, it was shown that exotic pairing between the bosons and the fermions might exist. However, the ground state phase diagram remained largely unknown, complicating any study to the influence of the experimentally unavoidable drawbacks, namely finite temperature and parabolic confinement. 
%The ground state phase diagram, which is the topic of this paper, is thus of interest to the whole field.
Using quantum Monte Carlo (QMC) simulations, we provide the complete, exact ground state phase diagram in one dimension at the parameters of greatest interest: double half filling with equal hopping $t_{\rm F}=t_{\rm B}$ . Afterwards, we discuss the case of strong anisotropic hopping $t_{\rm F}=4t_{\rm B}$. Recently, the grand-canonical QMC simulations of Ref.~\cite{Sengupta05} focused on the issue of the stability of the mixed phase.\\

We assume that a mixture of bosons and fermions is contained in a square well and loaded into an optical lattice. 
The temperature is low enough such that quantum degeneracy is achieved. 
The system is then described by a lowest-band Bose-Fermi Hubbard chain,
\begin{eqnarray}
H & = & - \sum_{\langle i, j \rangle}^L\left(  t_{\rm B} b^{\dagger}_ib_j + t_{\rm F} c^{\dagger}_ic_j  + {\textrm {h.c.}} \right) \nonumber \\
{} & {} & + \sum_i^L \frac{U_{\rm BB}}{2} n_i(n_i-1) + \sum_i^L U_{\rm BF} n_im_i,\label{eq:bfh}
\end{eqnarray}
where $b_i$ and $c_i$ are the bosonic and fermionic annihilation operators on site $i$, respectively. The intersite spacing $a$ is set to unity, i.e., $a=1$. The operators $n_i=b^{\dagger}_ib_i$ and $m_i=c^{\dagger}_ic_i$ denote the number operators on site $i$ for bosons and fermions, respectively. Bosons (fermions) can hop from site $i$ to a nearest neighbor site $j = i \pm 1$ with tunneling amplitude $t_{\rm B}$ ($t_{\rm F}$). Furthermore, a large occupation of bosons on a single site is suppressed by the on-site repulsion term $U_{\rm BB}$. Bosons and fermions can mutually repel or attract each other on every site depending on the sign of $U_{\rm BF}$. In order to enhance the pairing correlations in the system, we assume that bosons and fermions both have a density corresponding to double half filling, $N_{\rm B} = N_{\rm F} = L/2$. We work in units $t_{\rm B}=1$, and simulate at inverse temperature $\beta = 2L$. In a homogeneous system at half fermionic filling, the particle-hole transformation, $c_i \to (-1)^i c_i^{\dagger}$ changes the sign of the boson-fermion interaction in eq.(\ref{eq:bfh}), $U_{\rm BF} \to -U_{\rm BF}$. We choose $U_{\rm BF} > 0$ without loss of generality.\\

%Under the particle-hole transformation, $c_i \to (-1)^i c_i^{\dagger}$, the boson-fermion interaction term in  Hamiltonian (\ref{eq:bfh}) changes sign, $U_{\rm BF} \to -U_{\rm BF}$ while the other terms do not change. At fermionic half filling the physics is thus the same irrespective of the sign of $U_{\rm BF}$. In a parabolic trap, however, the above symmetry is lost. \\

One-dimensional phases can be characterized by the slowest decaying modes of the system. The bosonic Green function $G_{\rm B}(i-j) = \langle b_i^{\dagger}b_j \rangle$, fermionic Green function $G_{\rm F}(i- j)= \langle c_i^{\dagger}c_j \rangle$ and composite pair Green function $G_{\rm BF}(i-j) = \langle b_i^{\dagger}c_i^{\dagger}b_jc_j \rangle$ ($U_{\rm BF}<0$) or $G_{\rm BF}(i-j) = \langle b_i^{\dagger}c_j^{\dagger}b_jc_i \rangle $ ($U_{\rm BF}>0$)  allow for such a classification. 
In addition, the bosonic superfluid density $\rho_{\rm s,B} = W_{\rm B}^2L/(2 \beta)$~\cite{Pollock87}, the fermionic stiffness $\rho_{\rm s,F}= W_{\rm F}^2L/(2 \beta)$, the paired superfluid density $\rho_{\rm PSF} = (W_{\rm B} + W_{\rm F})^2L/(2 \beta)$~\cite{Kagan02} and the counter-rotating superfluid density $\rho_{\rm SCF}= (W_{\rm B} - W_{\rm F})^2L/(2 \beta)$~\cite{Kuklov04} probe for soft modes in the system. The winding numbers for the bosons and the fermions are denoted by $W_{\rm B}$ and $W_{\rm F}$, respectively.
The Green functions can either decay exponentially or by power law according to (for a continuum coordinate $x \gg 1$):
\begin{eqnarray}
G_{\rm B}(x) \sim  (d(x \vert L))^{-K_{\rm B}}, \label{eq:bosdecay} \\
G_{\rm F}(x) \sim \sin( \pi x/2) d(x \vert L)^{-K_{\rm F}}, \label{eq:ferdecay}\\
G_{\rm BF}(x) \sim \sin( \pi x/2) d(x \vert L)^{-K_{\rm BF}}, \label{eq:pairdecay}
\end{eqnarray}
where the cord function $d(x \vert L)  = \vert \sin( \pi x/L) \vert$ takes care of the periodic boundaries~\cite{Cazalilla04}. 
The relevant correlators for density-density correlations are $C_{\rm CDW(SDW)} = \langle (n_i \pm m_i)(n_j \pm m_j)\rangle - \langle (n_i \pm m_i)\rangle \langle (n_j \pm m_j)\rangle \sim \cos(\pi x) d(x \vert L)^{-K_{\rm CDW(SDW)}}$, where the upper signs refer to the charge-density wave correlator (CDW) and the lower signs to the spin-density wave (SDW) correlator.\\%They map onto each other when $U_{\rm BF} \to - U_{\rm BF}$.\\
%We can thus directly extract the Luttinger parameters from the simulations.\\

We study the Bose-Fermi Hubbard model using a quantum Monte Carlo method based on worm-type updates~\cite{Prokofev98}, using a generalization of the canonical methods of Refs.~\cite{Rombouts05, Vanhoucke05}. 
In a molecular superfluid phase bosons and fermions are paired into molecules and thus have to wind together. 
The standard single-particle worm algorithm changes the winding of the species independently. In order to create a molecular winding number, it needs to go through a state with different bosonic and fermionic winding numbers, which is exponentially suppressed. A key ingredient is the use of a two-particle, boson-fermion worm~\cite{Pollet05}, which can let the bosons and fermions move together. Technical details will be presented elsewhere.\\

\begin{figure}
\includegraphics[scale=0.6]{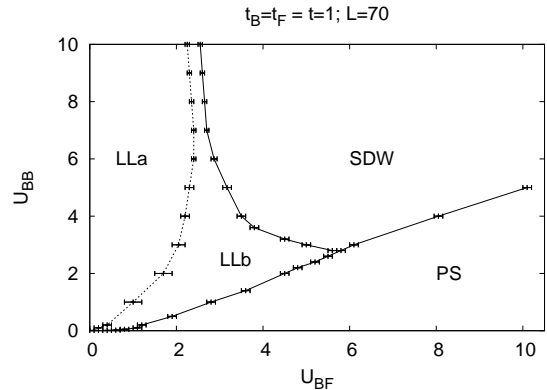}
\caption{\label{fig:phasediagram} Phase diagram of the Bose-Fermi Hubbard model, Eq.(~\ref{eq:bfh}). The different phases are (i) Phase separation ('PS'), (ii) Luttinger liquid phase ('LLa' and 'LLb'), and (iii) a (pseudo) spin-density wave ('SDW'). The phase transition (solid line) to the PS phase is first order, while the transition from LL to SDW belongs to the Kosterlitz-Thouless universality class~\cite{Kosterlitz73}. The dashed line indicates a cross-over when the composite pair Green function decays slower than the fermionic Green function. }
\end{figure}

\begin{figure}
\includegraphics[scale=0.5]{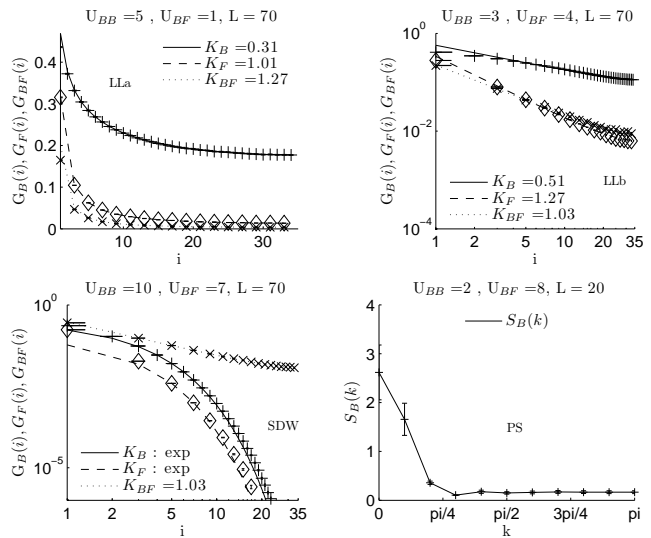}
\caption{\label{fig:green}Bosonic('+', full line), fermionic('$\diamond$', dashed line) and composite pair Green function('x', dotted line) in the phases LLa, LLb, SDW of Fig.~\ref{fig:phasediagram}. Only the exponential decay or power-law behavior in absolute value of the correlation functions has been shown. The PS phase has been identified by a peak in the bosonic structure factor at momentum $k = 2\pi/L$, shown in the fourth plot. }
\end{figure}

The main result of this study is the phase diagram of the Hamiltonian (\ref{eq:bfh}) at double half filling and with parameters $t_{\rm B}=t_{\rm F}=t$ shown in Fig.~\ref{fig:phasediagram}. We distinguish three phases:  (i) Luttinger liquid phase (LL), (ii) (pseudo) spin-density wave phase (SDW) and (iii) phase separation (PS). The Green functions in the various phases are shown in Fig.~\ref{fig:green}. We now make a tour clockwise around the phase diagram, starting at $U_{\rm BF} = 0$. \\

When there is no interaction between the bosons and the fermions, $U_{\rm BF}=0$, we have a system consisting of free fermions and a Luttinger liquid of bosons. For small $U_{\rm BF}$ the fermions also behave as a Luttinger liquid, which weakly interacts with the bosonic liquid. The system thus consists of two weakly interacting Luttinger liquids (LL phase). The bosonic Green function is always the slowest decaying mode. For small values of $U_{\rm BF}$, the fermionic Green function decays slower than the composite paired Green function (LLa), while there is a cross-over at larger $U_{\rm BF}$ when the fermionic Green function decays faster (LLb). \\

In the limit of infinite $U_{\rm BB}$, the bosons behave as hard-core or Tonks bosons~\cite{Girardeau60, Pollet04}). We can associate a pseudo-spin 'up' with the bosons and a pseudo-spin 'down' with the fermions. The charge sector becomes SU(2) symmetric and the system can be mapped onto a Heisenberg chain with effective spin-exchange amplitude $J \sim 4t^2/U_{\rm BF}$ for repulsive, small $U_{\rm BF}$. The Heisenberg chain has only one phase, a spin-density wave (SDW) with a gapless spin sector. There is a charge gap which is exponentially small, $\Delta_{\rm c} \sim \sqrt{U_{\rm BF}}e^{-v/U_{\rm BF}}$, with $v$ the velocity. In Fig.~\ref{fig:phasediagram} we see that the SDW phase extends to finite values of $U_{\rm BB}$. The phase is identified by exponential decay of the bosonic and the fermionic Green functions, shown in Fig.~\ref{fig:green}. The bosonic superfluid density and fermionic stiffness are small but non-zero, the pair superfluid density $\rho_{\rm PSF}$ is zero for repulsive $U_{\rm BF}$ while $\rho_{\rm SCF} = 2(\rho_{\rm s,B} + \rho_{\rm s,F})$. This is a reflection of the gapped charge and the gapless spin sector we identified in the limit $U_{\rm BB} \to +\infty$. The transition line between the LL and SDW phases converges logarithmically slowly to zero, as a consequence of the exponentially small charge gap.  In case of attractive $U_{\rm BF}$, the SDW phase  in Fig.~\ref{fig:phasediagram} is replaced by a singlet paired phase, where the power-law decay of the composite pair Green function is slower than the decay of the charge-density wave correlator. The $\rho_{\rm SCF}$ is zero and $\rho_{\rm PSF} = 2(\rho_{\rm s,B} + \rho_{\rm s,F})$. When $U_{\rm BB} \to \infty$ the system becomes isotropic and the composite pair Green function decays as fast as the charge-density wave correlator.\\

When both $U_{\rm BB}$ and $U_{\rm BF}$ are large, the model can be mapped onto an XXZ-Hamiltonian $H^{\rm XXZ} = \sum_i J(\sigma_i^x\sigma_{i+1}^{x} + \sigma_i^y\sigma_{i+1}^y) + J^z\sigma_i^z\sigma_{i+1}^z$ with $J = -\frac{t_{\rm B}t_{\rm F}}{U_{\rm BF}}$ and $J^z = \frac{t_{\rm B}^2 + t_{\rm F}^2}{2U_{\rm BF}} -\frac{t_{\rm B}^2}{U_{\rm BB}} $~\cite{Duan03}. Using the exact solution of the XXZ-chain, the first order transition towards phase separation (i.e., the ferromagnet) occurs at $J = J^z$ or $U_{\rm BF}/U_{\rm BB} = 2$ for $t_{\rm B}=t_{\rm F}=1$. This is the slope of the transition line between the SDW and PS phases in Fig.~\ref{fig:phasediagram}. \\

At infinite repulsion between the bosons and the fermions, the bosons and the fermions phase separate (PS) and form non-overlapping domains. When $U_{\rm BB}=0$ and when $U_{\rm BF}$ is very large, the ideal Bose and Fermi gases are in first order unstable to phase separation with hard domain walls. The energy is minimized when the bosons occupy a number of $y=\pi (L/2)^{1/3}$ sites. This instability persists whenever $U_{\rm BF}/t > {\mathcal O}(L^{2/3})$ in first order when $U_{\rm BF} \to 0$.
%In the thermodynamic limit, pure soft core bosons interacting with fermions will thus always phase separate. 
At larger $U_{\rm BB}$, the boson repulsion exerts a pressure such that the region occupied by the bosons will grow.  
%In Fig.~\ref{fig:green} we see a clear peak at momentum $k = \frac{2\pi}{L}$ in the bosonic structure factor, which is the Fourier transform of the bosonic density-density correlation function. 
In Fig.~\ref{fig:green} we see a clear peak at momentum $k = \frac{2\pi}{L}$ in the bosonic structure factor.
In the simulations we detect phase separation by a diverging peak in the structure factor at small momenta, accompanied by an exponential decay of all three Green functions and zero winding numbers, $\rho_{\rm s,B} = \rho_{\rm s,F} = \rho_{\rm SCF} = \rho_{\rm PSF} = 0$.  The phase transition to the phase separated phase is strongly first order. \\

\begin{figure}
\includegraphics[scale=0.6]{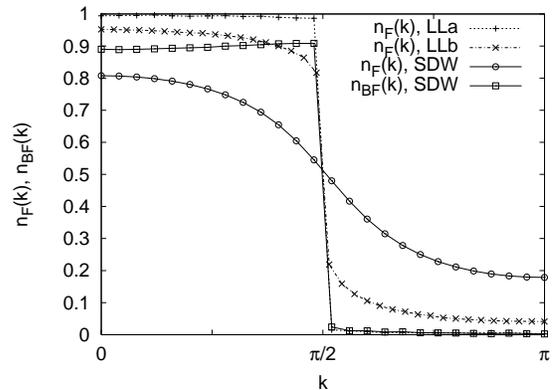}
\caption{\label{fig:momprof}Fermionic momentum profiles in the LLa, LLb and SDW phases for the same system parameters as in Fig.\ref{fig:green}. The error bars are smaller than the point sizes.
In the LL phase the Fermi momentum remains well defined, while the SDW phase destroys the Fermi surface of the bare fermions. The composite paired particles have a well defined Fermi momentum. 
}
\end{figure}

The physics of the different phases can also be understood when looking at the momentum profiles, which are experimentally accessible. In the LLa region, the fermionic momentum profile is very close to the non-interacting one. In the LLb, the fermionic momentum profile begins to deviate slightly, but the Fermi momentum is still well defined. In the SDW phase however, the fermions are strongly interacting while we see a Fermi surface for the paired composite particles. This is visualized in Fig.~\ref{fig:momprof}.\\

\begin{figure}
\includegraphics[scale=0.6]{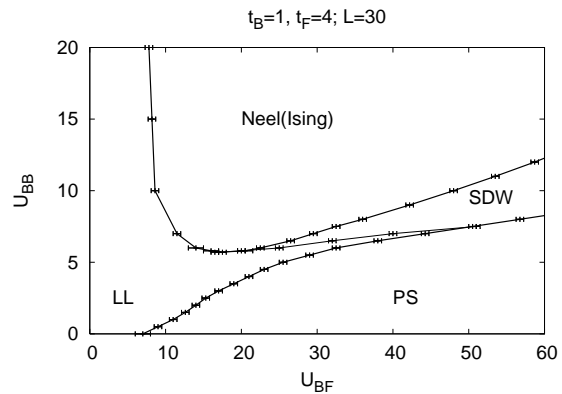}
\caption{\label{fig:phasediagram4} Phase diagram of the Bose-Fermi Hubbard model, Eq.(~\ref{eq:bfh}), at strong unequal hopping amplitudes. The different phases are (i) Phase separation ('PS'), (ii) Luttinger liquid phase ('LL'), (iii) a gapped 'Neel (Ising)' state with true long-range order and (iv) a gapless (pseudo) spin-density wave ('SDW') without long-range order. 
}
\end{figure}

\begin{figure}
\includegraphics[scale=0.6]{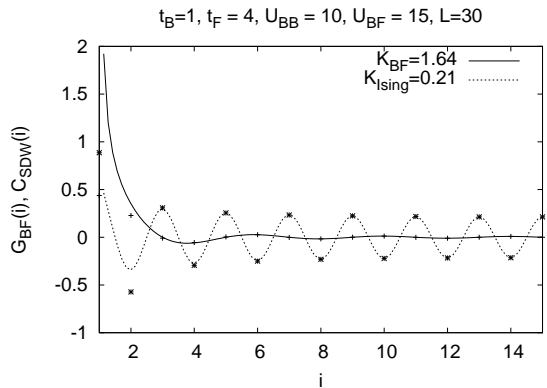}
\caption{\label{fig:structfact}Decay of the composite pair (boson-hole) Green function ($G_{\rm BF}$, '+' mark) and SDW order correlation function($C_{\rm SDW}$, '*' mark) in the Neel state. The data points are interpolated by power-law fits, see Eq.(\ref{eq:pairdecay}).} %We did not include the $1/x^2$ term in the SDW correlator such that the fit cannot go through the data point at $i=1$.}
\end{figure}

Difference in masses between the bosons and the fermions will in general lead to strong unequal hopping amplitudes~\cite{Cazalilla05}.
%True long-range order can only exist at (strong) unequal hopping amplitudes~\cite{Cazalilla05}. 
We therefore examine the system with parameters $t_{\rm F} = 4, t_{\rm B}=1$ at double half filling. At large $U_{\rm BB}$ and $U_{\rm BF}$ the XXZ Hamiltonian gives the asymptotic slope $U_{\rm BF}/U_{\rm BB} =4.5$ between a gapless phase and an anti-ferromagnetic phase (Neel phase), and for the transition to the ferromagnet (phase separation) the slope is $ U_{\rm BF}/U_{\rm BB} =12.5$. The full phase diagram is shown in Fig.~\ref{fig:phasediagram4}. The Neel (Ising) state is gapped and has true long-range order. The density-density part of the Ising correlator converges to a constant, as is shown in Fig.~\ref{fig:structfact}. Note that the phase transitions happen at much larger values of $U_{\rm BB}$ and $U_{\rm BF}$ compared to the case of equal hopping. In the gapless SDW phase, the composite pair boson-hole Green function has the slowest decay, while the pair superfluid density is zero, $\rho_{\rm PSF} = 0$. This is similar to the SDW state in Fig.~\ref{fig:phasediagram}. In case of Tonks bosons, $U_{\rm BB} \to \infty$, the transition from the LL phase to the Neel state belongs to the Kosterlitz-Thouless universality class~\cite{Fath95}.
The transition from PS to the Neel phase cannot occur directly, but a gapless phase exists between PS and the Neel phase.
The LL phase again has several crossovers, which we do not discuss here. In the limit $U_{\rm BB} = 0, U_{\rm BF} \to 0$ the same perturbation argument as in the equal hopping case would again lead to phase separation in the thermodynamic limit. From the present numerics on small system sizes it is difficult to confirm this, although it cannot be excluded either and we see an inflection point in the transition line between PS and LL at low values of $U_{\rm BF}$ in Fig.~\ref{fig:phasediagram4}.\\

The phases presented in this paper can experimentally be identified as follows. The momentum profiles of the bosons, fermions and composite particles distinguish the weakly interacting phases from phases characterized by strong pairing, as shown in Fig.~\ref{fig:momprof}. Noise correlations in time-of-flight images can be used to find the gapless SDW and gapped Ising/Neel phases~\cite{Greiner05, Foelling05}. In principle, RF spectroscopy can be used to measure the energy gaps~\cite{Chin04}. \\
%Currently, we are investigating other filling factors, such as the case of unit bosonic filling when a bosonic Mott phase can be formed. 
%How well the different phases can be resolved in a trapped system, is yet unknown, but simulations in a parabolic trap are within reach of the present algorithm and under way. \\

In conclusion, we have studied the ground state phase diagram of the Bose-Fermi Hubbard model at double half filling. Interactions can lead to such phases as a Luttinger liquid, a (pseudo) spin-density wave, a Neel(Ising) state and even phase separation can occur. \\

%In conclusion, we have studied the ground state phase diagram of the Bose-Fermi Hubbard model at double half filling. With equal tunneling amplitude $t_{\rm B}=t_{\rm F}$, the system is either a Luttinger liquid or a (pseudo) spin-density wave for repulsive $U_{\rm BF}$ (singlet paired state for attractive $U_{\rm BF}$). In the latter phase the bosons and the fermions pair into fermionic molecules, as shown in Fig.~\ref{fig:momprof}, and they are superfluid. Moreover, for $U_{\rm BF} \ge 2 U_{\rm BB}$ an instability towards phase separation occurs. 
%When the fermions are much more mobile than the bosons, $t_{\rm F} \gg t_{\rm B}$, the system develops a charge-density wave with true long-range order for attractive $U_{\rm BF}$. For repulsive $U_{\rm BF}$ the system is in a Neel state. The transition from the Neel state to phase separation occurs via a gapless state in between. 

We are grateful to T. Esslinger, T. Giamarchi, H. G. Katzgraber, C. Kollath, L. Pryadko and P. Sengupta for useful discussions. %, and P. Dayal for help with the IETL Lanczos solver. 
LP is financially supported by the Swiss National Science Foundation, KVH and SR by the Fonds voor Wetenschappelijk Onderzoek (FWO), Flanders. Simulations ran on the Hreidar cluster at ETH Z{\"u}rich. 

%\bibliography{bfh_re}

\end{document}